\documentstyle[aps,epsf,epsfig]{revtex}
\begin{document}
\twocolumn[\hsize\textwidth\columnwidth\hsize\csname @twocolumnfalse\endcsname 
\title{Implications of reflectance measurements on the mechanism for
superconductivity in MgB$_2$}
\author{F. Marsiglio}

\address{Department of Physics, University of Alberta, Edmonton,
Alberta, Canada T6G 2J1}

\date{\today} 
\maketitle 
\begin{abstract} 
Recent optical studies in c-axis oriented superconducting MgB$_2$ films
indicate that the electron-phonon coupling is weak \cite{tu01}. We 
reinforce this conclusion by examining the raw reflectance data; its 
frequency dependence is incompatible with strong electron-phonon scattering.
This is further strengthened by analysis of the real part of the conductivity,
and by the temperature dependence of the effective Drude scattering rate.
Using a realistic electron-phonon spectral shape \cite{kong01}, we find 
$\lambda_{\rm tr} \approx 0.15$, in agreement with Tu {\it et al.}
\cite{tu01}. To the extent that $\lambda_{\rm tr} \approx \lambda$, this
disagrees sharply with model calculations \cite{kong01,kortus01,an01},
and is far too low to provide the means for $T_c = 39 $ \, K. A simple model
is constructed with coupling to a high frequency excitation, which is
consistent with both the low frequency optical data and the high $T_c$. 
\end{abstract}

\vskip2pc]
 
The discovery of superconductivity in MgB$_2$ at 39 K \cite{nagamatsu01} 
has naturally prompted questions about mechanism. Thus far calculations
of the electron-phonon coupling strength \cite{kong01,kortus01,an01}
suggest that the conventional electron-phonon mechanism may be 
responsible for $T_c = 39$ K. Nonetheless, there remains some concern
that a significant modification of the theory is required to 
explain the high $T_c$, due to anharmonic \cite{yildrim01} or
non-adiabatic \cite{alexandrov01,cappelluti01} effects. In addition,
non-electron-phonon mechanisms have been proposed \cite{hirsch01,imada01}.

All of these deliberations have been essentially theoretical in nature.
On the experimental side, the isotope effect measurements \cite{budko01,hinks01}
can be understood within the electron-phonon framework, but only with
very high frequency phonons, and a very large Coulomb repulsion 
\cite{hinks01,knigavko01}. Very recently reflectance measurements on
c-axis oriented thin films have been analyzed within the conventional
electron-phonon framework \cite{tu01}. These authors conclude that the
electron-phonon coupling in MgB$_2$ is very weak, $\lambda_{tr} = 0.13 \pm
0.02$. The purpose of this paper is to further analyze their data, with
the goal of exploring possible ways in which the reflectance data can
be interpreted with a more sizeable electron-phonon interaction. We find,
in agreement with Tu {\it et al.} \cite{tu01} that the reflectance data
are {\bf not consistent} with an electron-phonon interaction sufficiently
strong to produce $T_c = 39$ K. Consistency can be achieved if another
excitation is assumed to exist at much higher frequency.

Through Kramers-Kronig (KK) analysis of the reflectance data, 
Tu {\it et al.} \cite{tu01} obtain a very Drude-like response at low
frequencies. As noted previously for Ba$_{1-x}$K$_x$BiO$_3$ (BKBO) 
\cite{dolgov91,marsiglio95,marsiglio96} this will 
be the case even if the electron-phonon
interaction is sizeable. In fact, the low frequency Drude result for the
real part of the conductivity is given by (see Eq. (13) in \cite{marsiglio95})
\begin{equation}
\sigma_{Drude} \equiv {\omega_{\rm Peff}^2 \over 4 \pi}
\Biggl({1/\tau_{\rm eff} \over \nu^2 + (1/\tau_{\rm eff})^2}\Biggr),
\label{drude}
\end{equation} 
where, at low temperature, 
$1/\tau_{\rm eff} = 1/\tau_{\rm imp}/(1 + \lambda_{\rm tr})$, where
$1/\tau_{\rm imp}$ is the elastic scattering rate due to impurities and 
$\lambda_{\rm tr}$ is the mass enhancement parameter defined specifically
for transport \cite{scher70,allen71}. Similarly $\omega_{\rm Peff} = 
\omega_{\rm P}/\sqrt{1 + \lambda_{\rm tr}}$ is the effective plasma frequency,
with $\omega_{\rm P}$ the bare plasma frequency.
This expression properly reduces to the correct zero frequency limit,
where renormalization effects are known to cancel \cite{prange64}. In what
follows we make no distinction between the conventional mass enhancement 
factor $\lambda$ and the transport version, $\lambda_{\rm tr}$, although the
latter is to be used in the conductivity while the former enters the single electron
self energy \cite{scher70,allen71}.  We will return to this point later. 
The experimentally derived parameters are
determined in the normal state at T = 45 K \cite{tu01}:
$\omega_{\rm Peff} = 13 600$ \, cm$^{-1}$, and $1/\tau_{\rm eff}
 = 75$ \, cm$^{-1}$. Any model with higher energy excitations that interact with
the electrons (such as phonons) must correctly reproduce this low 
frequency behaviour.

To attempt to model this data within the conventional electron-phonon
framework we compute the normal state self-energy \cite{allen82} and
compute the optical conductivity effectively in the bubble approximation 
\cite{allen71,marsiglio01}, since vertex corrections are
implicitly contained in the use of $\lambda_{\rm tr}$ rather
than $\lambda$. Calculation of the complex conductivity allows one to
compute all the optical functions, and in particular the reflectance
\cite{timusk89}.

In Fig. 1 we show the reflectance as a function of frequency, for a variety
of model calculations with different overall electron-phonon coupling 
strengths. For this purpose we have used the
model electron-phonon spectrum of Kong {\it et al.} \cite{kong01} for the
spectral function, $\alpha^2F(\omega)$. As calculated, it has $\lambda = 0.85$,
but we have scaled it by a constant to achieve any desired coupling strength. 
It is characterized by one dominant peak at $\omega = 74$ meV (600 cm$^{-1}$),
which manifests itself in the reflectance as a relatively sharp downward
trend, the magnitude of which depends on the coupling strength. Even without
the KK analysis, then, one can see directly from
the reflectance data that the electron-phonon coupling in MgB$_2$ must be weak.
This statement is independent of the precise form of the spectral function
used, as we have verified with other model spectra. The key point is that
over the frequency range of the phonons (and a little beyond) the frequency-dependent
scattering rate increases in proportion to the coupling strength 
\cite{dolgov91,marsiglio97}, and this results in a reduced reflectance in this
frequency range. In this respect we should mention that the model calculation with
$\lambda = 0.15$ fits the data shown in Fig. 1 very well, but nonetheless fails
at higher frequency. Specifically the reflectance data \cite{tu01} is significantly
lower than the computed result, and shows that at higher frequencies other
aspects of the problem are required \cite{remark1}. 

Nonetheless, a model which
both fits the low frequency reflectance and is capable of producing $T_c = 39$ \,
K consists of a weak electron-phonon spectrum, along with a high frequency
excitation (of unknown origin) \cite{marsiglio89,kaufmann98}, as shown in Fig. 2.
For simplicity we have adopted a broad Lorentzian spectrum centred at $0.5$ eV.
The coupling strength to this additional peak is weak ($\lambda \approx 0.4$),
but, because it is at high frequency, the spectrum shown in Fig. 2 results in
$T_c = 39$ \, K (with $\mu^\ast(\omega_c = 5 eV) = 0.17$). With this spectrum
the optical reflectance is calculated and shown as the dotted curve in Fig. 1;
it is barely distinguishable from the previous result with $\lambda = 0.15$.
Note, however, that we have used $\lambda = 0.2$ for the scaled Kong spectrum,
because the high frequency Lorentzian contribution essentially renormalizes
the low frequency part: $\alpha^2_{\rm eff}F(\omega) \sim \alpha^2F(\omega)/
(1 + \lambda_{\rm Lor})$ \cite{marsiglio89}. Such a spectrum also improves the
agreement with the higher frequency reflectance data, but we do not attach
a great deal of significance to this fact since other factors also contribute
\cite{remark1}. We therefore have not made any attempts to optimize this agreement.

Note that the calculations shown in Fig. 1 all utilize different values of the
plasma frequency and the impurity scattering rate. For example, for the curve
calculated with $\lambda_{\rm tr} = 0.85$, we used $\omega_P = 13600 
\sqrt{1 + \lambda_{\rm tr}} = 18470$ \, cm$^{-1}$ and $1/\tau_{\rm imp} = 
75 (1 + \lambda_{\rm tr}) = 138 $ \, cm$^{-1}$. 

In Fig. 3 we illustrate the same point through the real part of the
optical conductivity. In Fig. (3a) we show the various models already discussed
in association with Fig. 1. In Fig. (3b) we focus on the `Holstein' 
sideband near and above the characteristic phonon energies. It is clear 
that a strong electron-phonon coupling (i.e. of order unity or more) disagrees
with the data.
Moreover, a two-band model would not immediately remedy this problem 
since this would lead to extra absorption, which would contradict the data.  
Also shown (dot-dashed curve) is the calculation with the spectrum of 
Fig. 2. Because of the renormalization of the low frequency spectrum
discussed above, the result agrees very closely with the weak coupling
calculation. 
 
Another way in which one can analyze the data is through the temperature
dependence of some of the optical functions. For this purpose we
focus on the temperature dependent scattering rate \cite{puchkov94,marsiglio95},
as defined by Eq. (14) of the latter reference. This scattering rate was
used to fit the low frequency optical conductivity with a Drude 
model at various temperatures \cite{tu01}. Two points determined experimentally
are indicated in Fig. 4. In addition we have plotted the results expected
for various electron-phonon coupling strengths, again using the Kong {\it et al.}
\cite{kong01} spectrum as a model for the spectral function. The data indicates
that the effective scattering rate more than doubles between 45 K and 295 K;
this shows that some inelastic scattering must be present. However, the
required coupling strength to reproduce this temperature dependence is
$\lambda_{\rm tr} \approx 0.15$. This is in agreement with our analysis
of the frequency-dependent reflectance, and agrees with the estimates of Tu
{\it et al.} based on the optical sum rule and the temperature dependent
DC resistivity. As they pointed out, this coupling strength is much too small
to account for electron-phonon driven superconductivity in this material.
Our {\it ad hoc} model of Fig. 2 again reproduces the correct behaviour
(dot-dashed curve), indicating that some unknown high frequency excitation 
may be responsible for superconductivity.

In summary we have further analyzed the frequency and temperature dependence 
of the optical reflectance and conductivity in c-axis oriented superconducting 
MgB$_2$ films. We have found very strong support for the conclusion of
Ref. \cite{tu01} --- namely the electron-phonon coupling strength as derived
from optical measurements is much too small to account for superconductivity
in MgB$_2$. In all, five different (but not independent) analyses have produced this
conclusion: 1) the discrepancy between the plasma frequency as determined by
the low frequency Drude fit to the conductivity and that determined by
the sum rule \cite{tu01}, 2) the fit to the temperature dependence of
the DC resistivity data \cite{tu01},
3) the low frequency dependence of the reflectance, 4) the magnitude of the
Holstein side band in the real part of the conductivity, and 5) the
temperature dependence of the Drude width of the low frequency conductivity.
These conclusions have been obtained through calculation with a realistic
electron-phonon spectral shape, obtained through calculation \cite{kong01}.
Nonetheless, they are quite general, and depend only on the fact that the
phonons in this material extend up to a maximum of 100 meV. The data of
Tu {\it et al.} \cite{tu01} shows very convincingly that the coupling 
to excitations up to these energies must be weak. 

In our opinion the most compelling test is the frequency dependence of the
reflectance, since this property emerges essentially as raw data, and
does not require any KK analysis. If we accept the possibility of a high
frequency excitation coupling to electrons to produce superconductivity,
as is modeled in Fig. 2, then other consequences would immediately follow.
For example all superconducting properties ought to be very BCS-like. 

However, a possiblility for
reconciling these results with electron-phonon driven superconductivity is
that $\lambda_{\rm tr}$ and $\lambda$ differ significantly. This would be
the case, if, for example, the electron-phonon scattering was dominated
by forward scattering \cite{kulic00}. In this case one should expect other 
unusual normal and superconducting state properties; this possibility should
be further explored, both experimentally and theoretically.

\acknowledgements
I would like to thank Jorge Hirsch for bringing these experimental
results to my attention, and for subsequent discussions.
This work was supported by the Natural Sciences and Engineering
Research Council (NSERC) of Canada and the Canadian Institute for Advanced
Research.

\begin{figure}[h]
\epsfig{figure=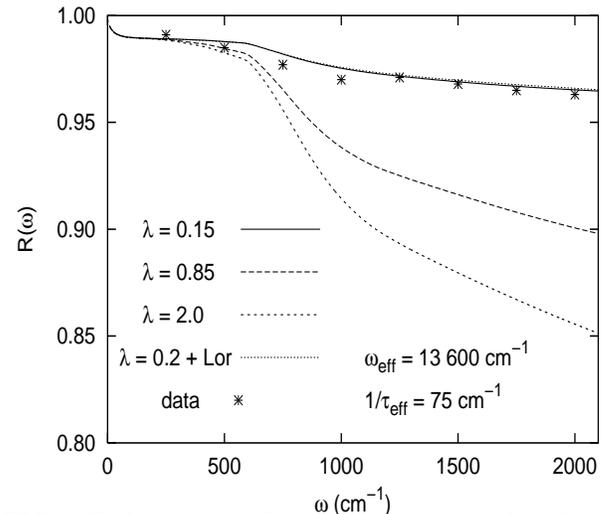,height=7cm,width=8cm}
\caption {Reflectance vs. frequency; experimental results are shown
with symbols, and curves indicate calculations, taking
into account different amounts of electron-phonon scattering. The
plasma frequency and the impurity scattering rate are determined for
each coupling strength to yield the effective plasma frequency and the
effective scattering rate indicated. The
data is in clear agreement with very weak electron-phonon coupling,
including the case where a higher frequency excitation also couples to the
electrons.
}
\label{Fig1}
\end{figure}

\begin{figure}
\epsfig{figure=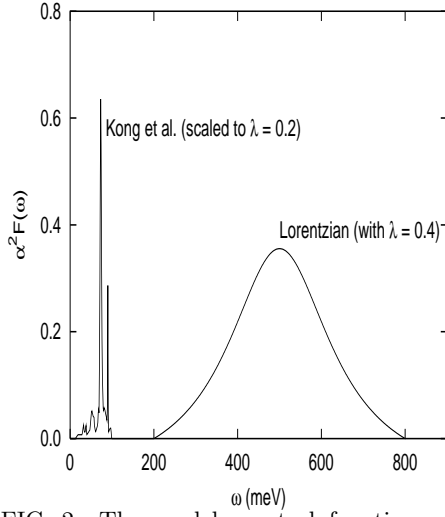,height=7cm,width=6cm}
\caption {The model spectral function, consisting of two pieces: the
electron-phonon part, taken from Kong {\it et al.} \protect\cite{kong01}
and scaled to give $\lambda = 0.2$, and a Lorentzian shaped contribution
peaked at 0.5 eV, and of unknown origin.
}
\label{Fig2}
\end{figure}

\begin{figure}
\epsfig{figure=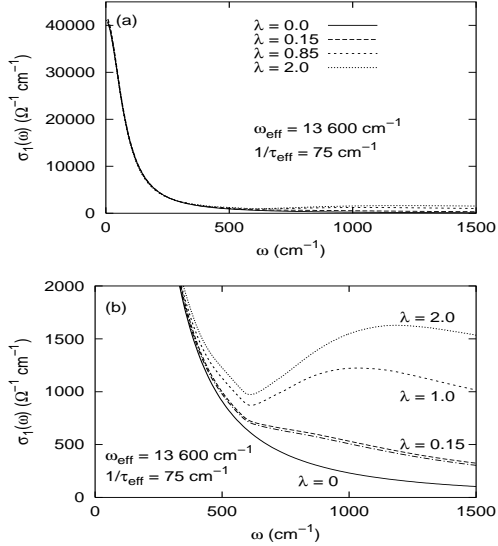,height=9cm,width=8cm}
\caption {Real part of the optical conductivity vs. frequency, (a) showing
the Drude-like peak that has been measured in thin film samples, and (b)
on a finer scale, emphasizing the Holstein absorption due to the electron-phonon 
coupling.  As in Fig. 1, the plasma frequency and impurity scattering rate
were determined separately for each coupling strength to yield the
observed low frequency Drude-like peak. The data (not shown) are in much better
agreement with the weak electron-phonon coupling scenario.
}
\label{Fig3}
\end{figure}   

\begin{figure}
\epsfig{figure=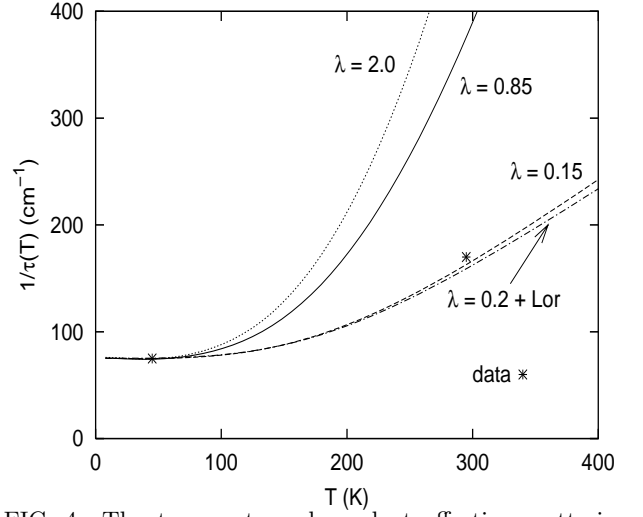,height=7.cm,width=8cm}
\caption {The temperature dependent effective scattering rate, determined
by the amount of electron-phonon coupling. The two data points reported
are indicated with symbols. They also indicate that the electron-phonon
coupling is weak.
}
\label{Fig4}
\end{figure}


\begin{references}
\bibitem{remark1} For example, it is clear from other data that other bands are
relevant at these energy scales, and anisotropic scattering also complicates the
analysis.
\end{references}
\end{document}